# Semiclassical thermoelectric transport in disordered Dirac electron system $Ag_2Te$

Kentaro Kuga,[1,2,3,4,*] Keisuke Hirata,[1,3] Daiki Goto,[1] Ryogo Ishihara,[1] Masaharu Matsunami,[1,2,3,5] and Tsunehiro Takeuchi[1,2,3,5,6]

[1]Toyota Technological Institute, Nagoya, Aichi 468-8511, Japan
[2]CREST, Japan Science and Technology Agency, Chiyoda-ku, Tokyo 102-0076, Japan
[3]MIRAI, Japan Science and Technology Agency, Chiyoda-ku, Tokyo 102-0076, Japan
[4]Graduate School of Science and Technology, Hirosaki University, Hirosaki, Aomori, 036-8561, Japan.
[5]Research Center for Smart Energy Technology, Toyota Technological Institute, Nagoya, Aichi 468-8511, Japan
[6]Institute of Innovation for Future Society, Nagoya University, Nagoya, Aichi 464-8603, Japan

**ABSTRACT**. We investigated the thermoelectric effects of the Dirac electron system $Ag_2Te$ under magnetic field. Our analysis based on the Boltzmann semiclassical model associated the disorder with the unconventional magnetic field responses such as linear magnetoresistance, linear Nernst effect, step-like Nernst effect, and sign change in Nernst effect. The analysis also revealed the impurity band near the Fermi energy. We simultaneously clarified the serious impact of the thermal Hall effect on the measurement of the Nernst effect, and we proposed the definitive solution. Our careful measurement and analysis will be the standard for the thermoelectric study under magnetic field.

Heat current transfers charges and is bended by magnetic field, generating a transverse electric field. This is the Nernst effect and is a thermoelectric analogue of the Hall effect. The Nernst effect is applicable to practical uses such as power generation and heat sensor [1–3], and has advantages over the Seebeck effect in higher efficiency and simpler structure in device [4,5]. Nernst effect is mainly categorized into normal Nernst effect described by Boltzmann semiclassical model and anomalous Nernst effect induced by Berry curvature. Recently, anomalous Nernst effect in magnetic Weyl semimetals has been extensively studied because of the interest in both basic and applied physics [6–9]. Some nonmagnetic Dirac electron systems are also reported to show the anomalous Nernst effect as the step-like magnetic field dependence which cannot be explained by the simple Boltzmann semiclassical model [10–14].

Normal Nernst effect is one of few ways to probe the energy dependence of the relaxation time $\partial\tau/\partial\varepsilon|_{\varepsilon=\xi}$ at the chemical potential $\xi$ although single-carrier system is required [15,16]. Using the Nernst effect, $\partial\tau/\partial\varepsilon|_{\varepsilon=\xi}$ in heavy fermion systems is discussed to confirm the local Kondo scattering which contributes to the characteristic phenomenon such as the anomalous Hall effect and the high-performance thermoelectric property [16–20]. Because $\tau$ in heavy fermion systems is caused by the sharp and large density of state of the $4f$ electrons, we expected that the impurity band also arises the considerable $\partial\tau/\partial\varepsilon|_{\varepsilon=\xi}$ and the Nernst effect. Impurity band has a sharp density of state but is difficult to probe because of the tiny intensity. Therefore, we investigated the Nernst effect in the Dirac electron system $Ag_2Te$ where strong effect of disorder plays role in the physical properties [21–28]. Normal Nernst effect is theoretically understood as follows.

Thermoelectric motive force originating from the temperature-dependent carrier diffusion induces charge current represented by the product of the temperature gradient $-\vec{\nabla}T$ and the thermoelectric conductivity tensor $\bar{\alpha}$, similarly to the product of the electric field $\vec{E}$ and the electrical conductivity $\bar{\sigma}$. If there are both $-\vec{\nabla}T$ and $\vec{E}$, the electrical current density $\vec{J}$ will be $\vec{J}=\bar{\sigma}\vec{E}-\bar{\alpha}\vec{\nabla}T$. This situation is illustrated in Fig. 1. Assuming the open circuit $\vec{J}=0$, the zero transverse temperature gradient $\partial T/\partial y=0$, and the isotropic physical properties, Seebeck effect $S_{xx}$ and Nernst effect $S_{xy}$ are described as

$$S_{xx}=\alpha_{xx}\rho_{xx}+\alpha_{xy}\rho_{yx} \tag{1}$$

$$S_{xy}=\alpha_{xy}\rho_{xx}-\alpha_{xx}\rho_{yx}, \tag{2}$$

where $\alpha_{xx}$, $\alpha_{xy}$, $\rho_{xx}$, and $\rho_{yx}$ are the Peltier conductivity, the Nernst conductivity, the electrical resistivity, and the Hall resistivity, respectively. Eqs. (1) and (2) are mathematically derived in Supplemental Material (SM) [29] including the transverse temperature gradient $\partial T/\partial y\neq 0$ for the practical experiment. We note that $S_{xx}$ and $S_{xy}$ assume $\partial T/\partial y=0$.

In single-band systems described by the simple Boltzmann semiclassical model within the relaxation time approximation, the magnetic field dependences of the Peltier conductivity $\alpha_{xx}^{\mathrm{Bol}}$ and the Nernst conductivity $\alpha_{xy}^{\mathrm{Bol}}$ are described as

$$\alpha_{xx}^{\mathrm{Bol}} = \frac{S_0\sigma_0}{1 + (\mu_{\mathrm{N}}B)^2} \tag{3}$$

$$\alpha_{xy}^{\mathrm{Bol}} = \frac{S_0\sigma_0\mu_{\mathrm{N}}B}{1 + (\mu_{\mathrm{N}}B)^2}, \tag{4}$$

where $S_0$, $\sigma_0$, $\mu_{\mathrm{N}}$, and $B$ are the zero-field Seebeck coefficient, the zero-field electrical conductivity, the Nernst mobility, and the magnetic field, respectively. Normally, the magnetic field dependences of $\rho_{xx}$ and $\rho_{yx}$ are $\rho_{xx} \cong \rho_0$ (constant) and $\rho_{yx} = \mu_{\mathrm{H}}\rho_0 B$, where $\rho_0$ and $\mu_{\mathrm{H}}$ are the zero-field electrical resistivity and the Hall mobility, respectively. Equation (2) is then transformed into

$$S_{xy} \cong \frac{(\mu_{\mathrm{N}} - \mu_{\mathrm{H}})S_0 B}{1 + (\mu_{\mathrm{N}}B)^2}. \tag{5}$$

*Contact author: k.kuga@hirosaki-u.ac.jp

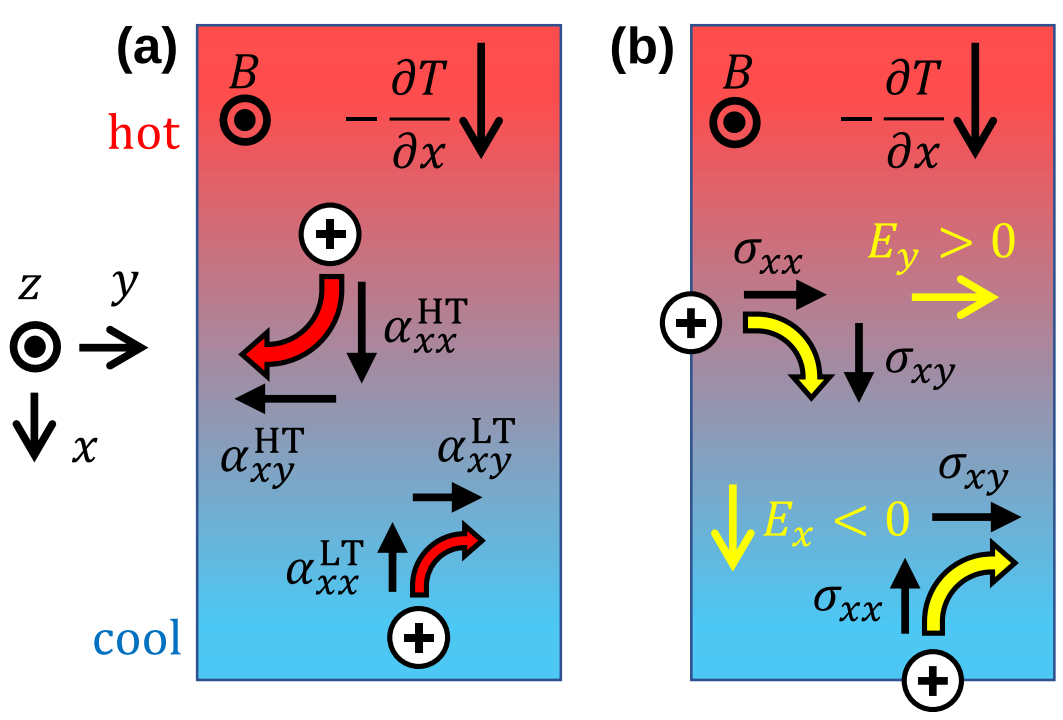

Fig. 1 Schematic drawing of the Seebeck effect and the Nernst effect in p-type material. (a) Carriers diffuse by heat and curve by the Lorentz force. Because the magnitude of diffusion is larger at the hotter side, thermoelectric motive forces are realized toward bottom and left sides which respectively correspond to the Peltier conductivity $\alpha_{xx} = \alpha_{xx}^{\mathrm{HT}} - \alpha_{xx}^{\mathrm{LT}}$ and the Nernst conductivity $\alpha_{xy} = \alpha_{xy}^{\mathrm{HT}} - \alpha_{xy}^{\mathrm{LT}}$. (b) In response to $\alpha_{xx}$ and $\alpha_{xy}$, carriers accumulate at the bottom and left sides, arising longitudinal and transverse electric fields $E_x$ and $E_y$. $E_x$ and $E_y$ yield the carrier flows in accordance with the electrical conductivity $\sigma_{xx}$ and the Hall conductivity $\sigma_{xy}$. (a) and (b) respectively represent $-\bar{\alpha}\vec{\nabla}T$ and $\bar{\sigma}\vec{E}$ which lead to $\vec{J} = \bar{\sigma}\vec{E} - \bar{\alpha}\vec{\nabla}T$.

At low enough temperatures where Mott relation is valid, $\mu_{\mathrm{N}} - \mu_{\mathrm{H}}$ is approximately transformed to be proportional to $\partial\tau/\partial\varepsilon|_{\varepsilon=\xi}$ [15]. Normally, $\partial\tau/\partial\varepsilon|_{\varepsilon=\xi}$ is negligibly small and $\mu_{\mathrm{N}} - \mu_{\mathrm{H}} \cong 0$, leading to tiny $S_{xy}$. This effect is called the Sondheimer cancellation [30].

$Ag_2Te$ forms a Monoclinic crystal structure with a space group of $P2_1/c$. Some Ag sites are partially occupied, which enables the easy introduction of disorder and the anharmonic lattice vibration, leading to a very low lattice thermal conductivity of 0.5 $\mathrm{Wm^{-1}K^{-1}}$ at room temperature [31,32]. A large and nonsaturating linear magnetoresistance even at room temperature is the most characteristic property which has been studied for more than two decades [21–27]. The unusual magnetoresistance is understood as a consequence of the network with various carrier mobilities due to the spatial disorder. Recently a large Nernst effect is also reported in off-stoichiometric $Ag_{1.9}Te$ especially around the temperatures where electron and hole compensate [33]. To avoid the complicated analysis of the Nernst effect, we selected the stoichiometric $Ag_2Te$ which has single-band properties [34].

Polycrystalline $Ag_2Te$ was prepared by the self-propagating high temperature synthesis method [32]. For the accurate measurement, we densified by the spark plasma sintering method at 350 °C and 50 MPa for 20 minutes, and annealed at 120 °C for 5 hours after which no change was confirmed in the thermoelectric properties. 98% of the theoretical density was confirmed by the Archimedes method. The sample was identified by powder XRD and EPMA, and we found no impurity phases and the stoichiometric composition.

Electrical and Hall resistivities were measured by the four-probe method using the resistivity option of Physical Property Measurement System (PPMS, Quantum Design, Inc.). Nernst effect and Seebeck effects were simultaneously measured by the steady-state method using chromel–constantan thermocouples. The temperature and the magnetic field were controlled by PPMS. The thermopowers of the thermocouples were converted into temperature differences from the base temperature by using the database provided by NIST [35]. The magnetic field was fixed during the thermoelectric measurement. The magnetic field effect in the thermocouple was corrected in accordance with the reference [36]. For eliminating the transverse signal in $S_{xx}$ and the longitudinal signal in $S_{xy}$ due to the misalignment in the electrical and thermal contact, we measured at both decreasing and increasing magnetic fields to extract the even component for $S_{xx}$ and the odd component for $S_{xy}$ (see SM for details [29]). Note that no hysteresis was observed before the extraction in all the measurements.

Figures 2 (e) and 2S (b) in SM [29] show the measurement setup and the thermocouples were attached not only along the longitudinal direction but also the transverse direction to monitor the transverse temperature difference $\Delta T_y$ induced by the thermal Hall effect. Because $\Delta T_y$ seriously impact on the experimental raw Nernst effect $S_{xy}^{\mathrm{raw}} = lV_y/w\Delta T_x$, we propose the solution described in SM [29]. Here, $l$, $w$, $V_y$, and $\Delta T_x$ are the sample length and width, the transverse voltage, and the longitudinal temperature difference, respectively. To justify the analysis, we compared between different experimental conditions of setups A and B illustrated in Fig. 2 (e) and (f), respectively. In setup A, the sample was sandwiched between Si plates at heater and heat sink sides for making a homogeneous heat flow. In setup B, a Si plate is anchoring to the sample to suppress $\Delta T_y$ [37]. In both setups, we used the same sample and Si plates were attached by using Torr Seal (Agilent Technologies, Inc.) which is an electrically insulating glue. We employed setup A for $S_{xx}$ and $S_{xy}$ shown in Fig. 3.

*Contact author: k.kuga@hirosaki-u.ac.jp

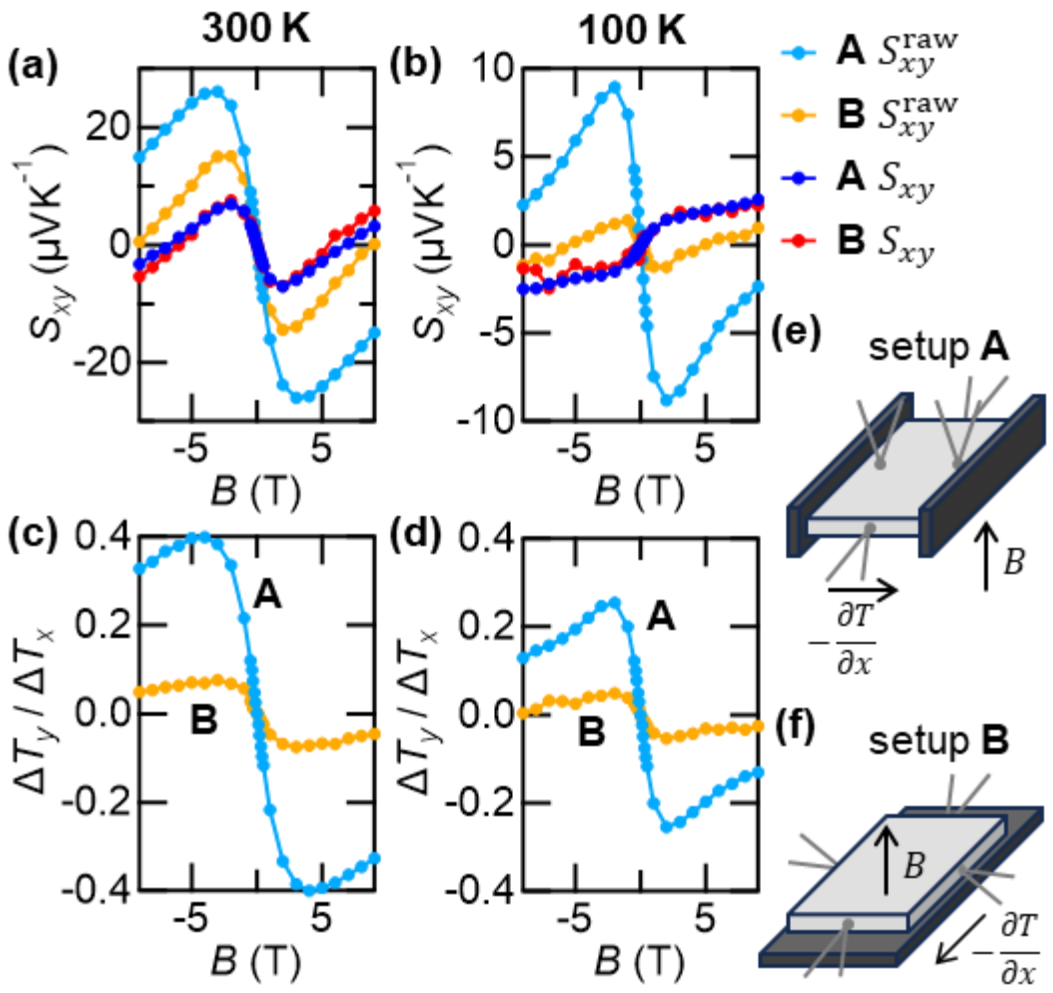

Fig. 2 Setup-dependent Nernst effect. Raw Nernst effect $S_{xy}^{\mathrm{raw}} = lV_y/w\Delta T_x$ is affected by the transverse temperature gradient $\Delta T_y$ induced by the thermal Hall effect, and intrinsic Nernst effect $S_{xy}$ is estimated by using Eqs. (S19) and (S20) in SM [29]. (a) and (c) are $S_{xy}^{\mathrm{raw}}$, $S_{xy}$, and $\Delta T_y/\Delta T_x$ at 300 K, and (b) and (d) are those at 100 K. (e) and (f) are the schematic illustration of the measurement setups A and B, respectively. Gray and black plates and silver lines indicate the samples, Si plates, thermocouples respectively. Same sample is used in both measurements. The directions of the measurements are 90° different within the plane of the original pellet, which should lead to the same physical properties confirmed by XRD [29].

So far, majority of the reports ignore the effect of $\Delta T_y$ on the measurement of the Nernst effect. Some reports employ a Si anchor similar to setup B to suppress $\Delta T_y$ [37], and some other reports measure $\Delta T_y$ to fix the result [38]. Here we introduce clear evidence of the serious impact by $\Delta T_y$ before showing the physical properties in $Ag_2Te$. Figures. 2 (a) and (b) indicate that both setups lead to completely different raw Nernst effect $S_{xy}^{\mathrm{raw}}$, and Figs. 2 (c) and (d) indicate that Si plate is not enough to suppress $\Delta T_y$ in spite of the thin sample with $t = 0.12$ mm. We then use Eqs. (S19) and (S20) in SM [29] to extract the intrinsic $S_{xy}$ as shown in Figs. 2 (a) and (b). $S_{xy}$ are completely different from $S_{xy}^{\mathrm{raw}}$ and even the sign changes are found at 100 K. However, $S_{xy}$ obtained by setups A and B are consistent, suggesting the validities of our measurement and analysis. These results indicate the importance of measuring $\Delta T_y$ to obtain the intrinsic $S_{xy}$. Theoretical influence of $\Delta T_y$ on the Nernst effect under the adiabatic condition and the situation where the impact is serious are discussed in SM [29]. We also confirmed the influence of $\Delta T_y$ on the anomalous Nernst effect in polycrystalline $Co_2MnGa$ [7] and the correction was 8% at 300 K as described in SM [29].

Figures 3 (a-d) show the magnetoresistance $MR$, $\rho_{yx}$, $S_{xx}$, and $S_{xy}$ at the temperatures from 10 to 300 K. $MR$ shows linear dependence above 1 T and parabolic-like dependence below 1 T at each temperature including 300 K, and is consistent with the previous report [21]. $B$-linear $\rho_{yx}$ is nearly temperature independent, suggesting the single-band contribution. The magnitude of $S_{xx}$ increases under magnetic fields at 300 K, similarly to the previous report [32]. $S_{xy}$ is much smaller than that in $Ag_{1.9}Te$ [33] because of the Sondheimer cancellation [30]. Below 60 K, we can find clear step-like magnetic field dependences which have been understood as the anomalous Nernst effect in some nonmagnetic topological materials [10–14].

For the detailed analysis, we extracted $\sigma_{xx}$, $\sigma_{xy}$, $\alpha_{xx}$, and $\alpha_{xy}$ by using $\sigma_{xx} = \rho_{xx}/(\rho_{xx}{}^2 + \rho_{yx}{}^2)$, $\sigma_{xy} = \rho_{yx}/(\rho_{xx}{}^2 + \rho_{yx}{}^2)$, and Eqs. (1) and (2) as shown in Figs. 3 (e-h). As a simple trial of the analyses of $\alpha_{xx}$ and $\alpha_{xy}$, we employed Eqs. (3) and (4) with the assumption of a negligible anomalous term and a variation of $\mu_{\mathrm{N}}$ because the variation of the mobility is used to explain linear $MR$ [21–24]. For the further simplicity, we assumed a Gaussian distribution in $\mu_{\mathrm{N}}$ and fixed $S_0$ and $\sigma_0$ as described in Eqs (S33) and (S34) in SMs [29]. The fittings using the assumptions above are also shown as the solid lines in Figs. 3 (g) and (h). The details of the analysis, the fitting without the variation of $\mu_{\mathrm{N}}$, and the fitting with an anomalous term are described in SM [29]. We concluded that the fittings in Figs. 3 (g) and (h) are the best among them because of the improvement in not only the fitting quality but also the consistency of the fitting parameters between $\alpha_{xx}$ and $\alpha_{xy}$ as shown in Figs. S5-S8 in SMs [29].

By employing experimental $\sigma_0$, the fitting parameters $S_0$ and $\mu_{\mathrm{N}}$ (average of the Nernst mobility) for $\alpha_{xx}$ and $\alpha_{xy}$ are obtained and plotted in Fig. 4 (a). $S_0$ is consistent with the experimental Seebeck coefficient at zero field (black line in Fig. 4(a)), suggesting that the analysis based on Eqs. (3) and (4) is proper. Therefore, $S_{xx}$ and $S_{xy}$ under magnetic field is governed by the Boltzmann semiclassical model and the anomalous Nernst effect is negligible. The analysis of the anomalous Nernst effect using the conventional empirical fitting [10–14] is also described in SM [29] and we again conclude the same statement. Similarly to $\alpha_{xx}$ and $\alpha_{xy}$, the fitting results in Figs. 3 (e) and (f) well reproduce $\sigma_{xx}$ and $\sigma_{xy}$ by introducing the variation of $\mu_{\mathrm{H}}$. The fitting parameter of $\mu_{\mathrm{H}}$ (average of the Hall mobility) is shown in Fig. 4 (a) (see SM for details [29]). These results suggest that the unconventional electrical and thermoelectrical properties in $Ag_2Te$ are governed by the Boltzmann semiclassical model with the variation of the mobility.

*Contact author: k.kuga@hirosaki-u.ac.jp

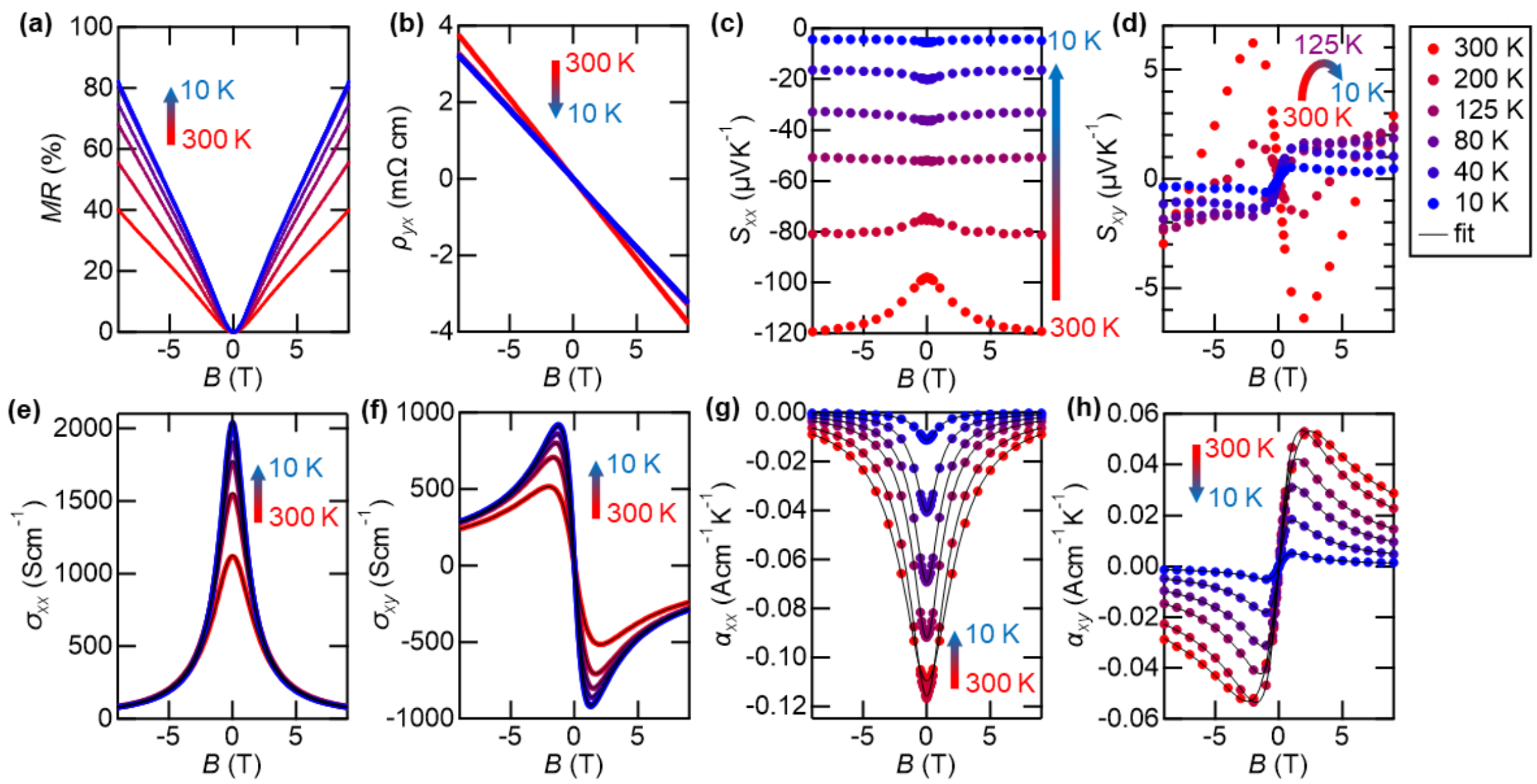

Fig. 3 Magnetic field dependences of the physical properties in $Ag_2Te$ at particular temperatures. Full data are shown in Figs. S5, S6, and S8 in SM [29]. (a) Magnetoresistance $MR$ represents $MR = \left(\rho_{xx}(B) - \rho_{xx}(0\ \mathrm{T})\right)/\rho_{xx}(0\ \mathrm{T})$. (b) Hall resistivities $\rho_{yx}$ overlap below 200 K. (c) and (d) $S_{xx}$ and $S_{xy}$ are already corrected by eliminating the effect of $\Delta T_y$ (see SM for details [29]). (e)–(f) Fitting curves are also shown.

Because Eqs. (1) and (2) include $\rho_{xx}$, large and linear $MR$ influences on the $S_{xx}$ and $S_{xy}$, resulting in step-like $S_{xy}$ as follows. According to Figs. 3 (a) and (b), $\rho_{xx}$ and $\rho_{yx}$ in $Ag_2Te$ are described as

$$\rho_{xx} = \rho_0(1 + MR) \tag{6}$$

$$\rho_{yx} = \mu_{\mathrm{H}}\rho_0 B. \tag{7}$$

Using Eqs. (1-4), (6), and (7), the magnetic field dependences of $S_{xx}$ and $S_{xy}$ are described as

$$S_{xx} = (1 + MR + \mu_{\mathrm{N}}\mu_{\mathrm{H}}B^2)\frac{S_0}{1 + (\mu_{\mathrm{N}}B)^2} \tag{8}$$

$$S_{xy} = \left((\mu_{\mathrm{N}} - \mu_{\mathrm{H}}) + \mu_{\mathrm{N}}MR\right)\frac{S_0 B}{1 + (\mu_{\mathrm{N}}B)^2}. \tag{9}$$

Here, fixed $\mu_{\mathrm{H}}$ and $\mu_{\mathrm{N}}$ are used for the simplicity, which is justified at low enough temperatures because the fittings of $\alpha_{xx}$ and $\alpha_{xy}$ using fixed $\mu_{\mathrm{N}}$ are acceptably match with the experiment as shown in the inset of Figs. S6 (a) and (c) in SM [29]. Eq. (9) indicates a constant value at the high magnetic field limit and suggests that linear $MR$ is responsible for the step-like magnetic field dependence in $S_{xy}$ regardless the origin of linear $MR$. In Eq. (8), $MR$ has a small contribution because $MR$ is tiny at low magnetic fields and the third term is dominant at high magnetic fields. Therefore, $S_{xx}$ shows a nearly Lorentzian magnetic field dependence with the value of $S_0\mu_{\mathrm{H}}/\mu_{\mathrm{N}}$ at the high magnetic field limit.

Above 80 K, $S_{xy}$ shows nearly linear magnetic field dependences at high magnetic fields. Equation (9) suggests that $MR$ term which comes from $\alpha_{xy}\rho_{xx}$ is dominant because of large and linear $MR$. Figure S6 (c) in SM [29] shows larger deviations between experimental $\alpha_{xy}$ and the fittings using fixed $\mu_{\mathrm{N}}$ at higher temperatures, and the deviations are strongly suppressed by employing the variation of $\mu_{\mathrm{N}}$ as shown in Fig. S6 (d). This suggests that the variation of $\mu_{\mathrm{N}}$ is responsible for the nearly linear magnetic field dependence in $S_{xy}$.

At low magnetic fields, $\mu_{\mathrm{N}} - \mu_{\mathrm{H}}$ term of Eq. (9) is dominant if $\mu_{\mathrm{H}} \neq \mu_{\mathrm{N}}$. Evidentially, Fig. 4 (a) shows a

*Contact author: k.kuga@hirosaki-u.ac.jp

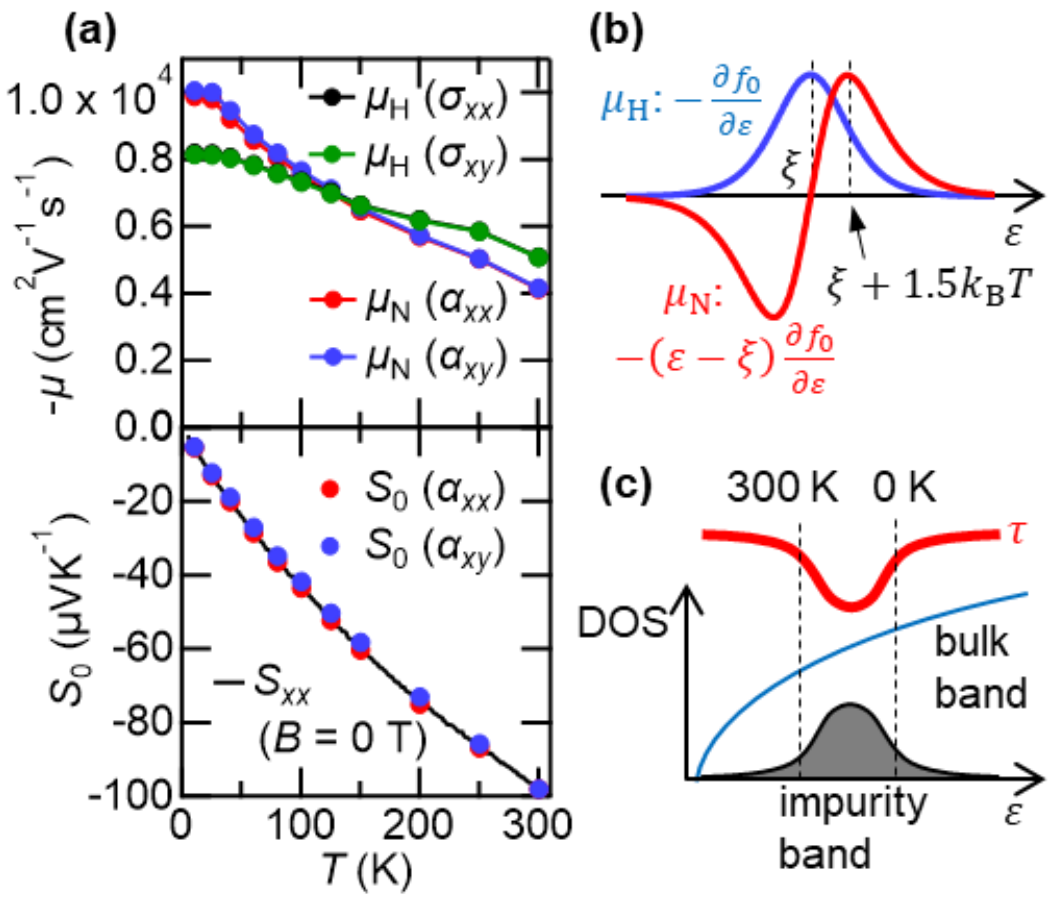

Fig. 4 (a) Fitting parameters of the Hall mobility $\mu_{\mathrm{H}}$, the Nernst mobility $\mu_{\mathrm{N}}$, and the zero-field Seebeck coefficient $S_0$. $\mu_{\mathrm{H}}$ estimated by $\sigma_{xx}$ and $\sigma_{xy}$ are overlapping. (b) Window functions of the Hall mobility $\mu_{\mathrm{H}}$ and the Nernst mobility $\mu_{\mathrm{N}}$. (c) Schematic illustration of the band structure and the relaxation time $\tau$. The broken lines represent the chemical potentials at 0 K and 300 K.

change in the magnitude relation between $\mu_{\mathrm{H}}$ and $\mu_{\mathrm{N}}$ at around 150 K. Figures. 3 (c) and (d) also show the sign change in the slopes of the magnetic field dependences in $S_{xx}$ and $S_{xy}$. According to the Boltzmann semiclassical model within the relaxation time approximation [39], $\mu_{\mathrm{H}}$ and $\mu_{\mathrm{N}}$ are described as

$$\mu_{\mathrm{H}} = \frac{-|e|}{m^*}\frac{\int D\tau^2 v^2 \left(-\frac{\partial f_0}{\partial \varepsilon}\right) d\varepsilon}{\int D\tau v^2 \left(-\frac{\partial f_0}{\partial \varepsilon}\right) d\varepsilon} \quad (10)$$

$$\mu_{\mathrm{N}} = \frac{-|e|}{m^*}\frac{\int (\varepsilon-\xi) D\tau^2 v^2 \left(-\frac{\partial f_0}{\partial \varepsilon}\right) d\varepsilon}{\int (\varepsilon-\xi) D\tau v^2 \left(-\frac{\partial f_0}{\partial \varepsilon}\right) d\varepsilon}, \quad (11)$$

where $e$, $m^*$, $\varepsilon$, $\xi$, $D$, $\tau$, $v$, and $f_0$ are the elementary charge, the electron effective mass, the energy of electron, the chemical potential, the density of state, the relaxation time, the group velocity, and the Fermi-Dirac distribution function, respectively. The denominators of the integral part of $\mu_{\mathrm{H}}$ and $\mu_{\mathrm{N}}$ are respectively the electrical conductivity and the Peltier conductivity, and are the normalization factor. The numerators are the relaxation times with the weights of $\varepsilon$-dependent conductivities. The $\varepsilon$-dependent conductivities are determined by the electronic structure with the weight of window functions which are schematically drawn in Fig. 4 (b). The window functions of $\mu_{\mathrm{H}}$ and $\mu_{\mathrm{N}}$ reflect the electronic structure near and $1.5k_{\mathrm{B}}T$ away from the chemical potential, respectively. Therefore, $\mu_{\mathrm{N}} - \mu_{\mathrm{H}}$ is roughly proportional to $\partial\tau/\partial\varepsilon|_{\varepsilon=\xi}$. At low enough temperatures where the Mott relation is valid, $\mu_{\mathrm{N}} - \mu_{\mathrm{H}}$ is theoretically transformed to be proportional to $\partial\tau/\partial\varepsilon|_{\varepsilon=\xi}$ [15]. This approximation would be roughly applicable at 300 K because the temperature dependence of $S_{xx}$ is roughly linear as shown in Fig. 4 (a). Note that the negative slope in the magnetic field dependence of $S_{xy}$ above 200 K cannot be realized by means of the hole excitation because it will be positive.

Because $Ag_2Te$ has an n-type carrier, $m^*$ is positive and positive $-(\mu_{\mathrm{N}} - \mu_{\mathrm{H}})$ leads to positive $\partial\tau/\partial\varepsilon|_{\varepsilon=\xi}$ in accordance with Eqs. (10) and (11). Figure 4 (a) then suggests that $\partial\tau/\partial\varepsilon|_{\varepsilon=\xi}$ is negative (positive) above (below) 150 K. In n-type materials, chemical potential decreases as increasing the temperature. Therefore, $\tau$ minimizes near the chemical potential at 150 K as schematically illustrated in Fig. 4 (c). Because the transport properties are strongly affected by disorder in $Ag_2Te$, the negative peak in $\tau$ would be the consequence of the scattering by an impurity band which density of state has a maximum near the chemical potential at 150 K.

Our thermoelectric measurement under the magnetic field linked disorder to the linear magnetic field dependences in $\rho_{xx}$ and $S_{xy}$, the step-like magnetic field dependence in $S_{xy}$, and the sign change in $S_{xy}$ near the zero magnetic field. Similar magnetic field responses are observed in some topological materials such as $Cd_3As_2$, $Bi_2Te_3$, $Bi_2Se_3$, and $Bi_{88}Sb_{12}$ [10,40–46]. Strong sample dependences are also reported. Therefore, the unconventional magnetic responses by the disorder observed in $Ag_2Te$ would be universal among high-mobility materials with disorder.

In summary, our careful thermoelectric measurements under magnetic field revealed the impact of the transverse temperature gradient on the Nernst effect and we proposed the definitive solution. The unconventional magnetic field responses of the linear magnetoresistance, the linear Nernst effect, the step-like Nernst effect, the sign change in the Nernst effect, and the enhancement of the Seebeck effect in single band $Ag_2Te$ could be comprehensively and quantitatively explained by the Boltzmann semiclassical model and the disorder. Our careful analysis probed that the step-like Nernst effect, which has been regarded as the anomalous Nernst effect in non-magnetic topological materials, can be accompanied by the linear magnetoresistance regardless the origin. The disordered state is revealed as the impurity band near the Fermi energy by means of the temperature dependent Nernst effect which probes the energy dependent relaxation time.

## ACKNOWLEDGMENTS

We thank T. Urata for fruitful discussion. This work was supported by JSPS KAKENHI (Grant Nos. 23K03327 and 26K08008) and JST CREST (Grant No.

*Contact author: k.kuga@hirosaki-u.ac.jp

JPMJCR18I2). The authors declare no conflicts of interest.

### DATA AVAILAVILITY

The data that support the findings of this study are available from the corresponding author upon reasonable request.

*Contact author: k.kuga@hirosaki-u.ac.jp

*Contact author: k.kuga@hirosaki-u.ac.jp

Supplemental Material for

# Semiclassical thermoelectric transport in disordered Dirac electron system $Ag_2Te$

Kentaro Kuga,[1,2,3,4] Keisuke Hirata,[1,3] Daiki Goto,[1] Ryogo Ishihara,[1] Masaharu Matsunami,[1,2,3,5] and Tsunehiro Takeuchi[1,2,3,5,6]

[1]Toyota Technological Institute, Nagoya, Aichi 468-8511, Japan
[2]CREST, Japan Science and Technology Agency, Chiyoda-ku, Tokyo 102-0076, Japan
[3]MIRAI, Japan Science and Technology Agency, Chiyoda-ku, Tokyo 102-0076, Japan
[4]Graduate School of Science and Technology, Hirosaki University, Hirosaki, Aomori, Japan.
[5]Research Center for Smart Energy Technology, Toyota Technological Institute, Nagoya, Aichi 468-8511, Japan
[6]Institute of Innovation for Future Society, Nagoya University, Nagoya, Aichi 464-8603, Japan

1. **Sign and subscript of physical quantities**

Because there are two different sign conventions [1] and the subscript inconsistency to describe the Nernst effect, we clearly define our description. When a two-dimensional physical quantity vector $\overrightarrow{A'}$ is a product of a tensor $\bar{T}$ and a two-dimensional vector $\vec{A}$, $\overrightarrow{A'}$ is described as

$$\overrightarrow{A'} = \begin{pmatrix} A'_x \\ A'_y \end{pmatrix} = \bar{T}\vec{A} = \begin{pmatrix} T_{xx} & T_{xy} \\ T_{yx} & T_{yy} \end{pmatrix} \begin{pmatrix} A_x \\ A_y \end{pmatrix}, \quad \text{(S1)}$$

where we omitted $z$ axis for the simplicity and employed the right-handed axes (Fig. S1 (a)). For example, the electrical and the Hall resistivities (resistivity tensor $\bar{\rho}$) are conventionally described as

$$\vec{E} = \begin{pmatrix} E_x \\ E_y \end{pmatrix} = \bar{\rho}\vec{J} = \begin{pmatrix} \rho_{xx} & \rho_{xy} \\ \rho_{yx} & \rho_{yy} \end{pmatrix} \begin{pmatrix} J_x \\ 0 \end{pmatrix} = \begin{pmatrix} \rho_{xx}J_x \\ \rho_{yx}J_x \end{pmatrix}, \quad \text{(S2)}$$

where $\vec{E}$ and $\vec{J}$ are the electric field vector and the electrical current density vector along $x$ axis, respectively. We then use "*yx*" in the subscript of the Hall resistivity. The thermoelectric coefficient tensor $\bar{S}$ is conventionally described as

$$\vec{E} = \begin{pmatrix} E_x \\ E_y \end{pmatrix} = \tilde{S}\vec{\nabla}T = \begin{pmatrix} S_{xx} & S_{xy} \\ S_{yx} & S_{yy} \end{pmatrix} \begin{pmatrix} \frac{\partial T}{\partial x} \\ 0 \end{pmatrix} = \begin{pmatrix} S_{xx}\frac{\partial T}{\partial x} \\ S_{yx}\frac{\partial T}{\partial x} \end{pmatrix}, \tag{S3}$$

where $\vec{E}$ and $\vec{\nabla}T$ are the electric field vector arisen by the thermoelectric motive force and the temperature gradient along $x$ axis, respectively, and the magnetic field is applied along $z$ axis (see Fig. S1 (b)). Because we conventionally describe Seebeck effect as $S_{xx} = E_x/(\partial T/\partial x)$, it is natural to describe Nernst effect as $S_{yx} = E_y/(\partial T/\partial x)$. $S_{yx}$ in Eq. (S3) would be then negative if the transverse electric field has the direction shown in Fig. 1 (b) in the main text. However, recently positive coefficient is widely used [1], so we use $S_{xy}$ $(= -S_{yx} = E_y/(-\partial T/\partial x)$ by the Onsager reciprocal relations) to follow the recent convention. As for the Hall and Nernst conductivities, we respectively use $\alpha_{xy}$ and $\sigma_{xy}$ because these descriptions represent positive in p-type systems and are frequently used.

*Contact author: k.kuga@hirosaki-u.ac.jp

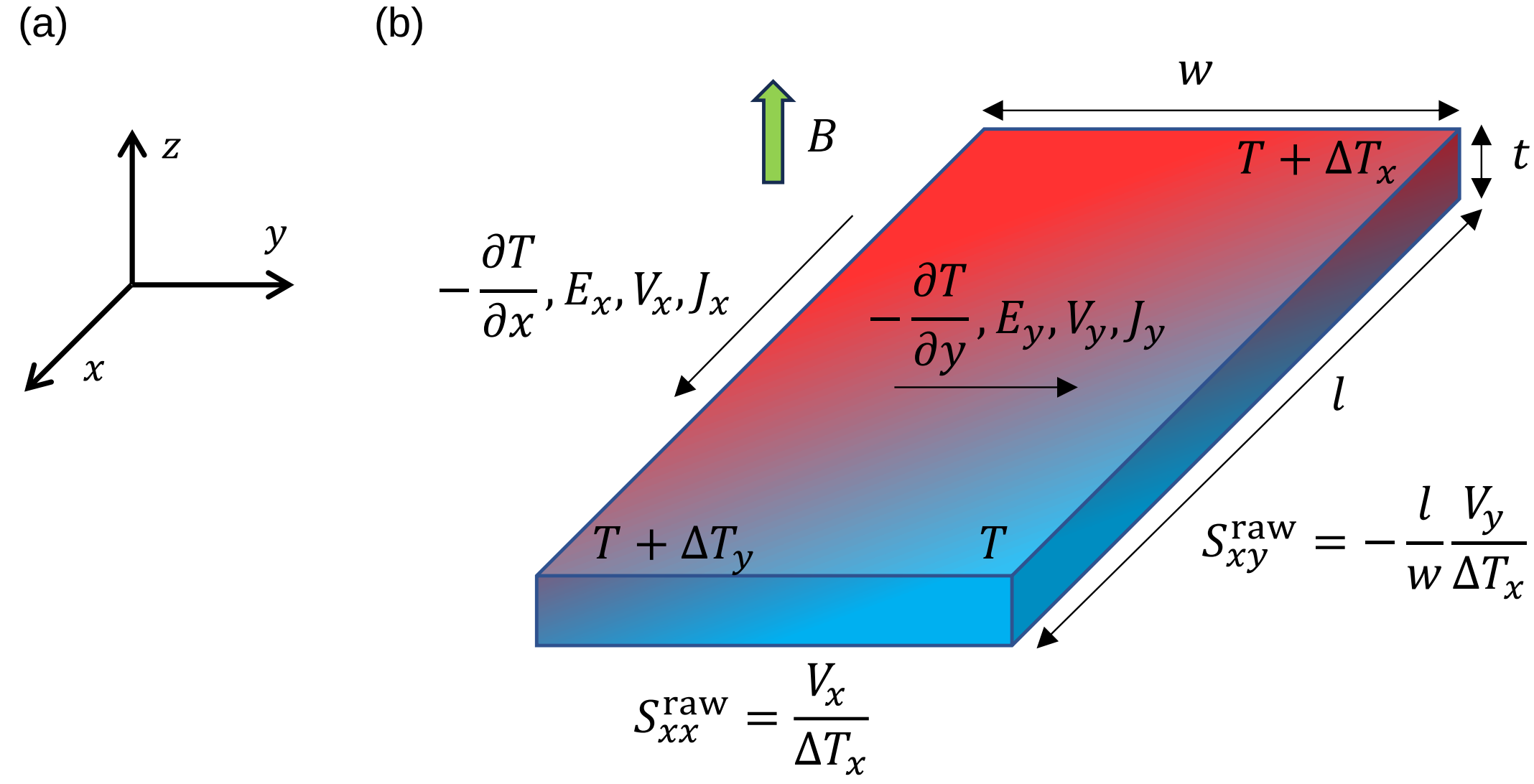

Fig. S1 Axis coordination (a) and the directions of physical quantities described in Section 1-3 (b).

## 2. **Correction of the measurement error due to misalignment**

In thermoelectric measurement, misalignment of the electrical and thermal contact at the sample can cause the contamination of the transverse ($y$-axis) signals in the longitudinal ($x$-axis) voltage $V_x$ and temperature difference $\Delta T_x$ and the longitudinal signals in the transverse voltage $V_y$ and temperature difference $\Delta T_y$. Because the longitudinal and transverse signals are respectively even and odd against magnetic field, the errors can be eliminated by extracting even or odd components in the magnetic field dependences. We then measured at both decreasing and increasing magnetic fields, and we eliminated the errors by using

$$V_x(B) = \frac{V_x^{\downarrow}(B) + V_x^{\uparrow}(-B)}{2} \tag{S4}$$

$$\Delta T_x(B) = \frac{\Delta T_x^{\downarrow}(B) + \Delta T_x^{\uparrow}(-B)}{2} \tag{S5}$$

$$V_y(B) = \frac{V_y^{\downarrow}(B) - V_y^{\uparrow}(-B)}{2} \tag{S6}$$

$$\Delta T_y(B) = \frac{\Delta T_y^{\downarrow}(B) - \Delta T_y^{\uparrow}(-B)}{2}, \tag{S7}$$

where the superscripts ↓ and ↑ indicate the measurement at decreasing and increasing magnetic field, respectively. Similar to the thermoelectric measurement, errors by the misalignment are also eliminated in the measurements of the electrical and Hall resistivities. Note that no hysteresis was observed in all the measurements.

*Contact author: k.kuga@hirosaki-u.ac.jp

3. **Equation between intrinsic thermoelectric properties and experiments**

In thermoelectric measurement under magnetic field, applied temperature gradient $\partial T/\partial x$ induces transverse temperature gradient $\partial T/\partial y$ by the thermal Hall effect. In this condition, we mathematically obtain the experimental Seebeck effect and the Nernst effect by using $\vec{J} = \bar{\sigma}\vec{E} - \bar{\alpha}\vec{\nabla}T$ with the open-circuit condition $\vec{J} = 0$. Here, $\bar{\sigma}$ and $\bar{\alpha}$ are the electric conductivity tensor and the thermoelectric conductivity tensor, respectively, and the transverse components of $\vec{E}$ and $\vec{\nabla}T$ are not zero ($E_y \neq 0$ and $\partial T/\partial y \neq 0$). This situation is illustrated in Fig. S1 (b) and we can transform to

$$\vec{J} = \bar{\sigma}\vec{E} - \bar{\alpha}\vec{\nabla}T = \begin{pmatrix} \sigma_{xx} & \sigma_{xy} \\ \sigma_{yx} & \sigma_{yy} \end{pmatrix} \begin{pmatrix} E_x \\ E_y \end{pmatrix} - \begin{pmatrix} \alpha_{xx} & \alpha_{xy} \\ \alpha_{yx} & \alpha_{yy} \end{pmatrix} \begin{pmatrix} \frac{\partial T}{\partial x} \\ \frac{\partial T}{\partial y} \end{pmatrix}$$

$$= \begin{pmatrix} \sigma_{xx}E_x + \sigma_{xy}E_y - \alpha_{xx}\frac{\partial T}{\partial x} - \alpha_{xy}\frac{\partial T}{\partial y} \\ -\sigma_{xy}E_x + \sigma_{yy}E_y + \alpha_{xy}\frac{\partial T}{\partial x} - \alpha_{yy}\frac{\partial T}{\partial y} \end{pmatrix}, \tag{S8}$$

*Contact author: k.kuga@hirosaki-u.ac.jp

where we used the Onsager reciprocal relations $\sigma_{yx} = -\sigma_{xy}$ and $\alpha_{yx} = -\alpha_{xy}$. With the open-circuit condition $\vec{J} = 0$ and isotropic conductivities $\sigma_{yy} = \sigma_{xx}$ and $\alpha_{yy} = \alpha_{xx}$ in polycrystalline samples, Eq. (S8) is transformed to

$$\sigma_{xx}E_x + \sigma_{xy}E_y - \alpha_{xx}\frac{\partial T}{\partial x} - \alpha_{xy}\frac{\partial T}{\partial y} = 0 \tag{S9}$$

$$-\sigma_{xy}E_x + \sigma_{xx}E_y + \alpha_{xy}\frac{\partial T}{\partial x} - \alpha_{xx}\frac{\partial T}{\partial y} = 0. \tag{S10}$$

Solving the simultaneous equations above, we obtain

$$E_x = \left(\rho_{xx}\alpha_{xx} + \rho_{yx}\alpha_{xy}\right)\frac{\partial T}{\partial x} + \left(\rho_{xx}\alpha_{xy} - \rho_{yx}\alpha_{xx}\right)\frac{\partial T}{\partial y} \tag{S11}$$

$$E_y = -\left(\rho_{xx}\alpha_{xy} - \rho_{yx}\alpha_{xx}\right)\frac{\partial T}{\partial x} + \left(\rho_{xx}\alpha_{xx} + \rho_{yx}\alpha_{xy}\right)\frac{\partial T}{\partial y}, \tag{S12}$$

where $\rho_{xx}$ and $\rho_{yx}$ are the electrical resistivity $\rho_{xx} = \sigma_{xx}/\left({\sigma_{xx}}^2 + {\sigma_{xy}}^2\right)$ and the Hall resistivity $\rho_{yx} = \sigma_{xy}/\left({\sigma_{xx}}^2 + {\sigma_{xy}}^2\right)$, respectively. Taking the volume integral, we obtain

$$-V_x wt = \left(\rho_{xx}\alpha_{xx} + \rho_{yx}\alpha_{xy}\right)wt(-\Delta T_x) + \left(\rho_{xx}\alpha_{xy} - \rho_{yx}\alpha_{xx}\right)lt\left(-\Delta T_y\right) \tag{S13}$$

$$-V_y lt = -\left(\rho_{xx}\alpha_{xy} - \rho_{yx}\alpha_{xx}\right)wt(-\Delta T_x) + \left(\rho_{xx}\alpha_{xx} + \rho_{yx}\alpha_{xy}\right)lt\left(-\Delta T_y\right), \tag{S14}$$

where $l$, $w$, and $t$ are respectively the length, width, and thickness of the sample as shown in Fig. S1 (b). We then obtain the experimentally raw Seebeck effect $S_{xx}^{\mathrm{raw}}$ and the Nernst effect $S_{xy}^{\mathrm{raw}}$:

$$S_{xx}^{\mathrm{raw}} = \frac{V_x}{\Delta T_x} = \left(\rho_{xx} - \rho_{yx}\tan\theta_{\mathrm{TH}}^{\mathrm{raw}}\right)\alpha_{xx} + \left(\rho_{yx} + \rho_{xx}\tan\theta_{\mathrm{TH}}^{\mathrm{raw}}\right)\alpha_{xy} \tag{S15}$$

$$S_{xy}^{\mathrm{raw}} = -\frac{l}{w}\frac{V_y}{\Delta T_x} = \left(\rho_{xx} - \rho_{yx}\tan\theta_{\mathrm{TH}}^{\mathrm{raw}}\right)\alpha_{xy} - \left(\rho_{yx} + \rho_{xx}\tan\theta_{\mathrm{TH}}^{\mathrm{raw}}\right)\alpha_{xx}, \tag{S16}$$

where we used $\tan\theta_{\mathrm{TH}}^{\mathrm{raw}} = l\Delta T_y/w\Delta T_x$ and $\theta_{\mathrm{TH}}^{\mathrm{raw}}$ is the experimental thermal Hall angle in any experimental conditions. Assigning $\Delta T_y = 0$ to Eqs. (S15) and (S16), we obtain the intrinsic (isothermal) Seebeck effect $S_{xx}$ and the Nernst effect $S_{xy}$:

$$S_{xx} = \alpha_{xx}\rho_{xx} + \alpha_{xy}\rho_{yx} \tag{S17}$$

$$S_{xy} = \alpha_{xy}\rho_{xx} - \alpha_{xx}\rho_{yx}. \tag{S18}$$

Equations (S17) and (S18) are the same as Eqs. (1) and (2) in the main text, respectively. Using Eqs (S15-18), we obtain

$$S_{xx} = \frac{S_{xx}^{\mathrm{raw}} - \tan\theta_{\mathrm{TH}}^{\mathrm{raw}}\,S_{xy}^{\mathrm{raw}}}{1 + \tan^2\theta_{\mathrm{TH}}^{\mathrm{raw}}} \tag{S19}$$

$$S_{xy} = \frac{S_{xy}^{\mathrm{raw}} + \tan\theta_{\mathrm{TH}}^{\mathrm{raw}}\,S_{xx}^{\mathrm{raw}}}{1 + \tan^2\theta_{\mathrm{TH}}^{\mathrm{raw}}}. \tag{S20}$$

*Contact author: k.kuga@hirosaki-u.ac.jp

We can evaluate intrinsic $S_{xx}$ and $S_{xy}$ using experimentally obtained $S_{xx}^{\mathrm{raw}}$, $S_{xy}^{\mathrm{raw}}$, $\Delta T_x$, $\Delta T_y$, $l$, $w$, and Eqs. (S19) and (S20).

4. **Estimation of extrinsic Seebeck effect and Nernst effect in measurement**

In the previous section, we introduced how to extract intrinsic $S_{xx}$ and $S_{xy}$ from the experimental results. In this section, we discuss when the extrinsic Seebeck and Nernst effects are considerable. Under the "adiabatic" condition, $\Delta T_y$ can be derived starting from

$$\overrightarrow{J^Q} = \bar{S}T\vec{J} - \bar{\kappa}\vec{\nabla}T, \tag{S21}$$

where $\overrightarrow{J^Q}$ and $\bar{\kappa}$ are the heat current vector and the thermal conductivity tensor, respectively. With the open-circuit situation $\vec{J} = 0$ and no heat leak toward transverse direction $J_y^Q = 0$, we obtain the relation between $\Delta T_y$ and $\tan\theta_{\mathrm{TH}} = \kappa_{xy}/\kappa_{xx}$:

$$\tan\theta_{\mathrm{TH}} = \frac{\kappa_{xy}}{\kappa_{xx}} = \frac{l\Delta T_y}{w\Delta T_x}, \tag{S22}$$

where $\kappa_{xx}$, $\kappa_{xy}$, and $\theta_{\mathrm{TH}}$ are the thermal conductivity, the thermal Hall conductivity, and the thermal Hall angle, respectively. Substituting experimental $S_{xx}^{\mathrm{raw}}$, $S_{xy}^{\mathrm{raw}}$, and $\theta_{\mathrm{TH}}^{\mathrm{raw}}$ for adiabatic $S_{xx}^{\mathrm{adi}}$, $S_{xy}^{\mathrm{adi}}$, and $\theta_{\mathrm{TH}}$, the simultaneous equations of Eqs. (S19) and (S20) lead to

$$S_{xx}^{\mathrm{adi}} = S_{xx} + S_{xy}\tan\theta_{\mathrm{TH}} \tag{S23}$$

$$S_{xy}^{\mathrm{adi}} = S_{xy} - S_{xx}\tan\theta_{\mathrm{TH}}\,. \tag{S24}$$

$\kappa_{xx}$ can be usually decomposed into the electron thermal conductivity $\kappa_{xx}^{\mathrm{ele}}$ and the lattice thermal conductivity $\kappa_{xx}^{\mathrm{lat}}$. The transverse lattice thermal conductivity is negligibly small. Assuming both longitudinal and transverse Wiedemann-Franz law, $\theta_{\mathrm{TH}}$ will be

$$\tan\theta_{\mathrm{TH}} = \frac{\kappa_{xy}}{\kappa_{xx}^{\mathrm{ele}} + \kappa_{xx}^{\mathrm{lat}}} = \tan\theta_{\mathrm{H}}\frac{1}{1 + \frac{\kappa_{xx}^{\mathrm{lat}}}{\kappa_{xx}^{\mathrm{ele}}}}. \tag{S25}$$

where $\theta_{\mathrm{H}}$ is the Hall angle. Equations (S24) and (S25) suggest that the extrinsic Nernst effect ($\sim -S_{xx}\tan\theta_{\mathrm{TH}}$) will be considerable when $S_{xx}$, $\tan\theta_{\mathrm{H}}$, and $\kappa_{xx}^{\mathrm{lat}}$ are respectively large, large, and small. These properties all match in $Ag_2Te$ and we need to correct this extrinsic effect by using Eqs. (S19) and (S20). Note that $-S_{xx}\tan\theta_{\mathrm{TH}}$ is ideally independent of the sample dimension and is normally negative at positive magnetic fields because the signs of $S_{xx}$ and $\tan\theta_{\mathrm{H}} = \sigma_{xy}/\sigma_{xx}$ are basically the same. Similar to the Nernst effect, the extrinsic Seebeck effect ($\sim S_{xy}\tan\theta_{\mathrm{TH}}$) will be considerable when $S_{xy}$, $\tan\theta_{\mathrm{H}}$, and $\kappa_{xx}^{\mathrm{lat}}$ are respectively large, large, and small. According to Eq. (S23), the relation $|S_{xy}| \ll |S_{xx}|$ leads to the insignificant extrinsic Seebeck effect in $Ag_2Te$.

*Contact author: k.kuga@hirosaki-u.ac.jp

5. **Influence of $\Delta T_y$ in other material**

We revealed the serious impact of $\Delta T_y$ in $Ag_2Te$ as shown in Fig. 2 in the main text. We here confirm the polycrystalline Weyl ferromagnet $Co_2MnGa$ at 300 K which has Giant anomalous Nernst effect [2]. Similarly to the majority of the reports of the Nernst effect, the setup A in Fig. 2 (c) in the main text is employed. Figure S2 shows $S_{xy}^{\mathrm{raw}}$, $S_{xy}$, and the picture of the measurement setup. $S_{xy}^{\mathrm{raw}}$ (= 6.4 μV/K at 1 T) is consistent with the previous report except the broader magnetic field dependence due to the demagnetizing effect. By using Eq. (S20), we extracted $S_{xy}$ (= 5.9 μV/K at 1 T) which is 8% smaller than $S_{xy}^{\mathrm{raw}}$. Because $Co_2MnGa$ has a large lattice thermal conductivity, the extrinsic Nernst effect is small as expected from Eq. (S24) and (S25) but is more or less significant. Here, the extrinsic Nernst effect in $Co_2MnGa$ is unusually positive at positive magnetic fields which is consistent with Eq. (S24) with negative $S_{xx}$ and positive $\sigma_{xy}$ [2]. The unusually different signs are realized because $S_{xx}$ and $\sigma_{xy}$ have different origins of normal thermoelectric effect and anomalous Hall effect, respectively.

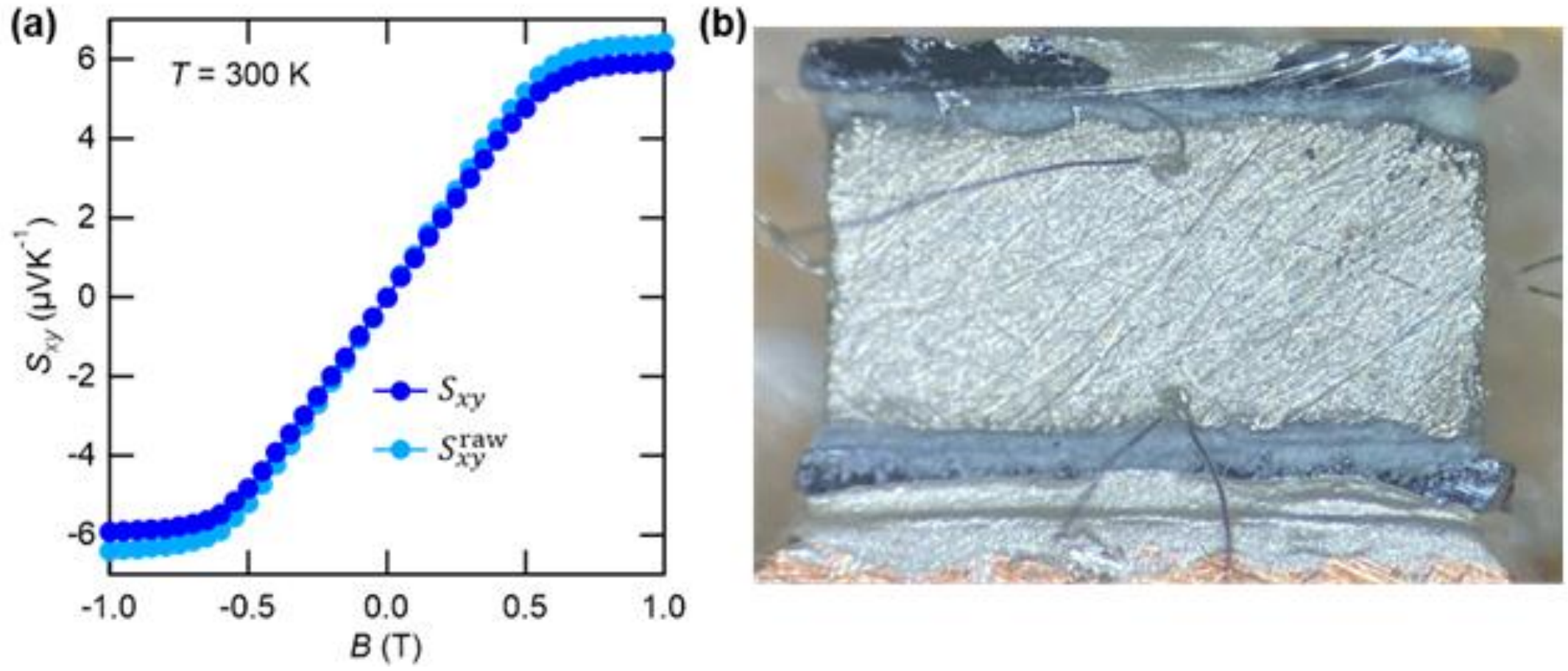

Fig. S2 Magnetic field dependence of $S_{xy}^{\mathrm{raw}}$ and $S_{xy}$ of $Co_2MnGa$ at 300 K (a) and the picture of the measurement setup (b).

6. **Sample identification and direction dependence**

After the synthesize described in the in the main text, the sample forms the pellet with the size of $\phi 6 \times 1$ mm$^3$. For identifying the crystal structure, we performed x-ray diffraction spectrum measurements with three different line scans parallel and perpendicular to the pressure in the spark plasma sintering process, and the results are shown in Fig. S3. The peak positions are consistent with the theoretical calculation of $Ag_2Te$ based on the previous report [3] and no extra peaks were observed. The relative intensities show a direction dependence with respect to the pressure direction but there are no differences within the plane of the pellet, suggesting the isotropic properties within the plane. Therefore, the relations of $\rho_{xx} = \rho_{yy}$, $S_{xx} = S_{yy}$, and $\kappa_{xx} = \kappa_{yy}$ are available.

*Contact author: k.kuga@hirosaki-u.ac.jp

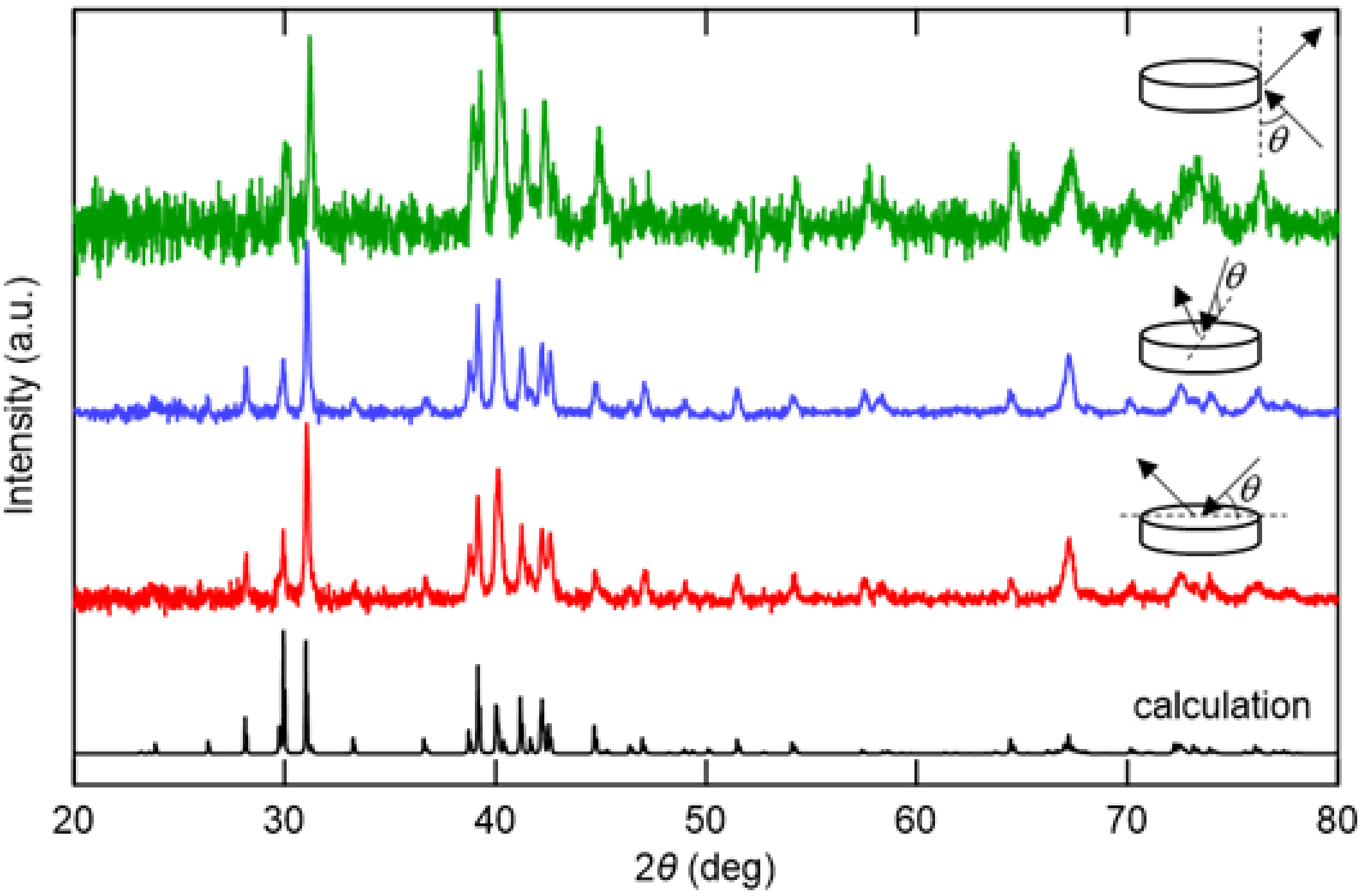

Fig. S3 X-ray diffraction patterns of the top and the side of the $Ag_2Te$ pellet.

## 7. Trial of *tanh* fitting and negligible anomalous Nernst effect

*tanh* fitting has been empirically used to confirm the existence of the anomalous Nernst effect in nonmagnetic topological materials [4–8]. We here try a similar fitting. For the quantitative analysis, we fitted $\alpha_{xy}\rho_{xx}$ ($1^{\mathrm{st}}$ term of $S_{xy}$ in Eq. S18) because this component in the single band system represents

$$\alpha_{xy}\rho_{xx} = \frac{S_0 \mu_{\mathrm{N}} B}{1 + (\mu_{\mathrm{N}} B)^2} + S_{xy}^{\mathrm{A}} \tanh \frac{B}{B_0}, \tag{S26}$$

where $S_{xy}^{\mathrm{A}}$ and $B_0$ are the amplitude of the anomalous Nernst effect and the saturation field, respectively. By fitting $\alpha_{xy}\rho_{xx}$, we can verify the consistency between $S_0$ and the experimental Seebeck coefficient $S_{xx}$ at the zero magnetic field. Figure S4 (a) shows the magnetic field dependence of $\alpha_{xy}\rho_{xx}$ and the fitting at each temperature. The fitting looks very well, but the fitting parameter $S_0$ is inconsistent with experimental $S_{xx}$ at the zero magnetic field while $S_0$ estimated by the fitting in Fig. 3 (h) in the main text is consistent. Therefore, the fitting of Eq. (S26) is inappropriate and the anomalous Nernst effect is negligibly small in $Ag_2Te$.

*Contact author: k.kuga@hirosaki-u.ac.jp

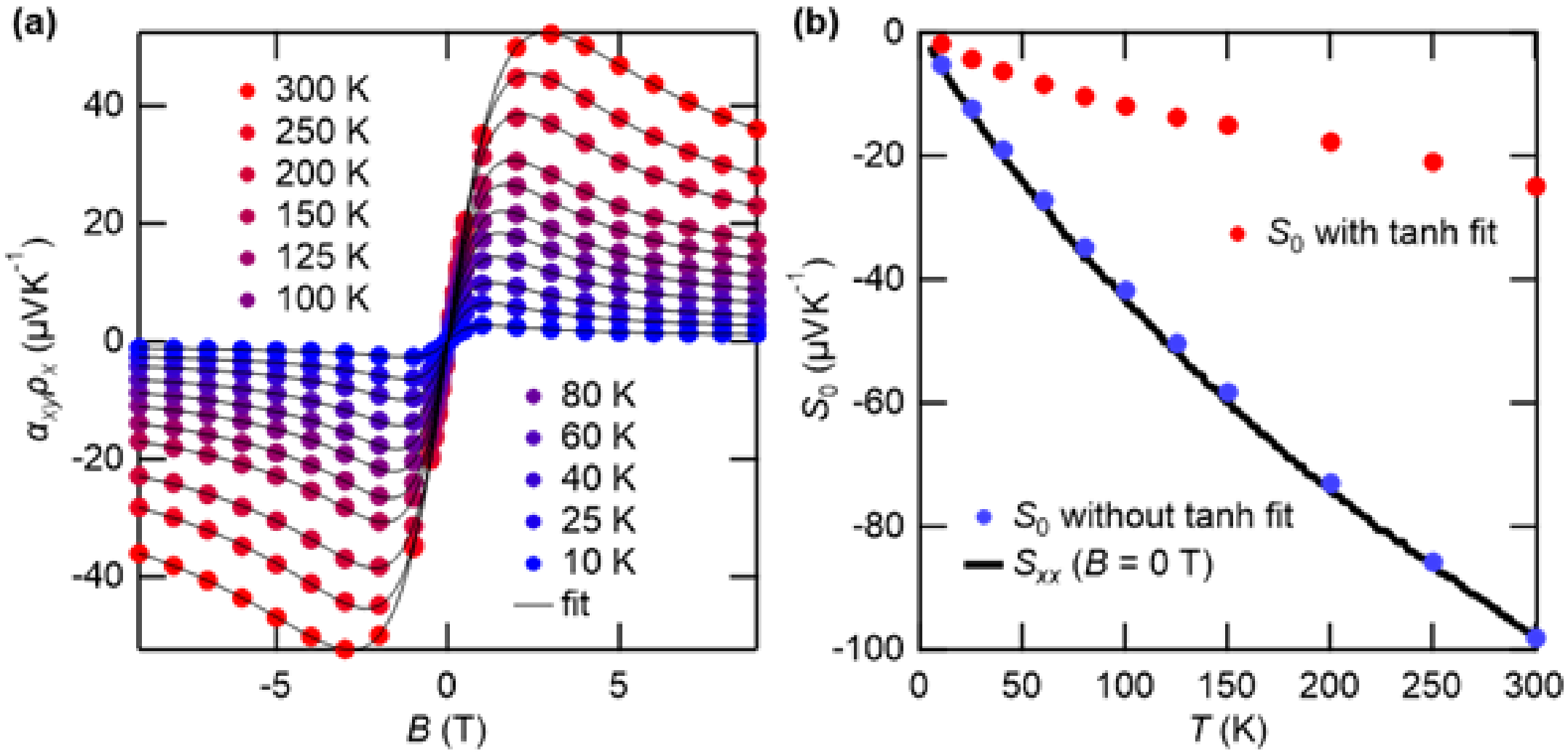

Fig. S4 Magnetic field dependence of $\alpha_{xy}\rho_{xx}$ with the fitting of Eq. (S26) at each temperature and the fitting parameter $S_0$. (b) shows the fitting parameter $S_0$ in (a) (with *tanh* fit), the one in Fig. 3 (h) in the main text (without *tanh* fit), and experimental $S_{xx}$ at zero field.

## 8. **Variation of Hall mobility and Nernst mobility**

In the Boltzmann semiclassical model, electrical conductivity $\sigma_{xx}^{\mathrm{Bol}}$, Hall conductivity $\sigma_{xy}^{\mathrm{Bol}}$, Peltier conductivity $\alpha_{xx}^{\mathrm{Bol}}$, and Nernst conductivity $\alpha_{xy}^{\mathrm{Bol}}$ is described as

$$\sigma_{xx}^{\mathrm{Bol}} = \frac{\sigma_0}{1+(\mu_{\mathrm{H}}B)^2} \tag{S27}$$

$$\sigma_{xy}^{\mathrm{Bol}} = \frac{\sigma_0\mu_{\mathrm{H}}B}{1+(\mu_{\mathrm{H}}B)^2} \tag{S28}$$

$$\alpha_{xx}^{\mathrm{Bol}} = \frac{S_0\sigma_0}{1+(\mu_{\mathrm{N}}B)^2} \tag{S29}$$

$$\alpha_{xy}^{\mathrm{Bol}} = \frac{S_0\sigma_0\mu_{\mathrm{N}}B}{1+(\mu_{\mathrm{N}}B)^2}. \tag{S30}$$

Because the variation of the mobility is used to explain linear *MR* [9–12], it is natural to take the variations of $\mu_{\mathrm{H}}$ and $\mu_{\mathrm{N}}$ into account. For the simplicity, we assumed the Gaussian distribution in $\mu_{\mathrm{H}}$ and $\mu_{\mathrm{N}}$ and fixed $\sigma_0$ and $S_0$. Then each conductivity $\sigma_{xx}^{\mathrm{Gau}}$, $\sigma_{xy}^{\mathrm{Gau}}$, $\alpha_{xx}^{\mathrm{Gau}}$, and $\alpha_{xy}^{\mathrm{Gau}}$ will be

*Contact author: k.kuga@hirosaki-u.ac.jp

$$\sigma_{xx}^{\text{Gau}} = \int \frac{\sigma_0}{1+(\mu B)^2} \frac{1}{\sqrt{2\pi}w} \exp\left(-\frac{(\mu-\mu_{\text{H}})^2}{2w^2}\right) d\mu \tag{S31}$$

$$\sigma_{xy}^{\text{Gau}} = \int \frac{\sigma_0 \mu B}{1+(\mu B)^2} \frac{1}{\sqrt{2\pi}w} \exp\left(-\frac{(\mu-\mu_{\text{H}})^2}{2w^2}\right) d\mu \tag{S32}$$

$$\alpha_{xx}^{\text{Gau}} = \int \frac{S_0 \sigma_0}{1+(\mu B)^2} \frac{1}{\sqrt{2\pi}w} \exp\left(-\frac{(\mu-\mu_{\text{N}})^2}{2w^2}\right) d\mu \tag{S33}$$

$$\alpha_{xy}^{\text{Gau}} = \int \frac{S_0 \sigma_0 \mu B}{1+(\mu B)^2} \frac{1}{\sqrt{2\pi}w} \exp\left(-\frac{(\mu-\mu_{\text{N}})^2}{2w^2}\right) d\mu\,, \tag{S34}$$

where $w$ is the peak width of the Gaussian distribution. In these equations, $\mu_{\text{H}}$ and $\mu_{\text{N}}$ represent the average of the Hall mobility and the Nernst mobility, respectively. In the integration, positive $\mu$ is cut off because $Ag_2Te$ is n-type. Figures S5 and S6 show experimental $\sigma_{xx}$, $\sigma_{xy}$, $\alpha_{xx}$, $\alpha_{xy}$, and their fittings of Eqs. (S27-S34). The data and the fittings in Figs. S5 (b) and (d) and Figs. S6 (b) and (d) are the same as those in Figs. 3 (e-h). Without the Gaussian distribution, the fittings deviate especially in $\sigma_{xx}$ because linear magnetoresistance cannot be reproduced by simple Boltzmann semiclassical model. The fittings of $\alpha_{xx}$ and $\alpha_{xy}$ at low temperatures acceptably match with the experimental results. By employing Gaussian distribution, we can find drastic improvements in the fittings. We then obtained $\mu_{\text{H}}$, $\mu_{\text{N}}$, and $S_0$ as shown in Fig. 4 (a) in the main text. The comparisons of the fitting parameters $\mu_{\text{H}}$ and $\mu_{\text{N}}$ between with and without Gaussian distribution are shown in Figs. S7 (a) and (b). We can find much better agreements between $\mu_{\text{H}}$ estimated by $\sigma_{xx}$ and $\sigma_{xy}$ and between $\mu_{\text{N}}$ estimated by $\alpha_{xx}$ and $\alpha_{xy}$, suggesting the justification of our analysis. The normalized variations $\Delta\mu/\mu = w/\mu_{\text{H}}$ or $w/\mu_{\text{N}}$ are shown in Fig. S7 (c) and all $\Delta\mu/\mu$ are roughly consistent. We also reconstructed $S_{xx}$ and $S_{xy}$ by using experimental $\rho_{xx}$, $\rho_{xy}$, and the fitting parameters of $S_0$, $\mu_{\text{N}}$, and $w$. For comparison, both calculations with (using Eqs. (S33), (S34), and the fittings in Figs. S6 (b) and (d)) and without (using Eqs. (S29), (S30), and the fittings in Figs. S6 (a) and (c)) Gaussian distribution are shown in Fig. S8. We can find drastic improvements in both $S_{xx}$ and $S_{xy}$ by introducing Gaussian distribution. However, the calculations with Gaussian distribution are still slightly different from the experimental results in spite of the quantitative agreements in our analysis, suggesting the slightly imperfect assumption of our model. Note that the difference in $S_{xy}$ looks larger than that in $S_{xx}$ because the vertical axis range is about 10 times narrower.

*Contact author: k.kuga@hirosaki-u.ac.jp

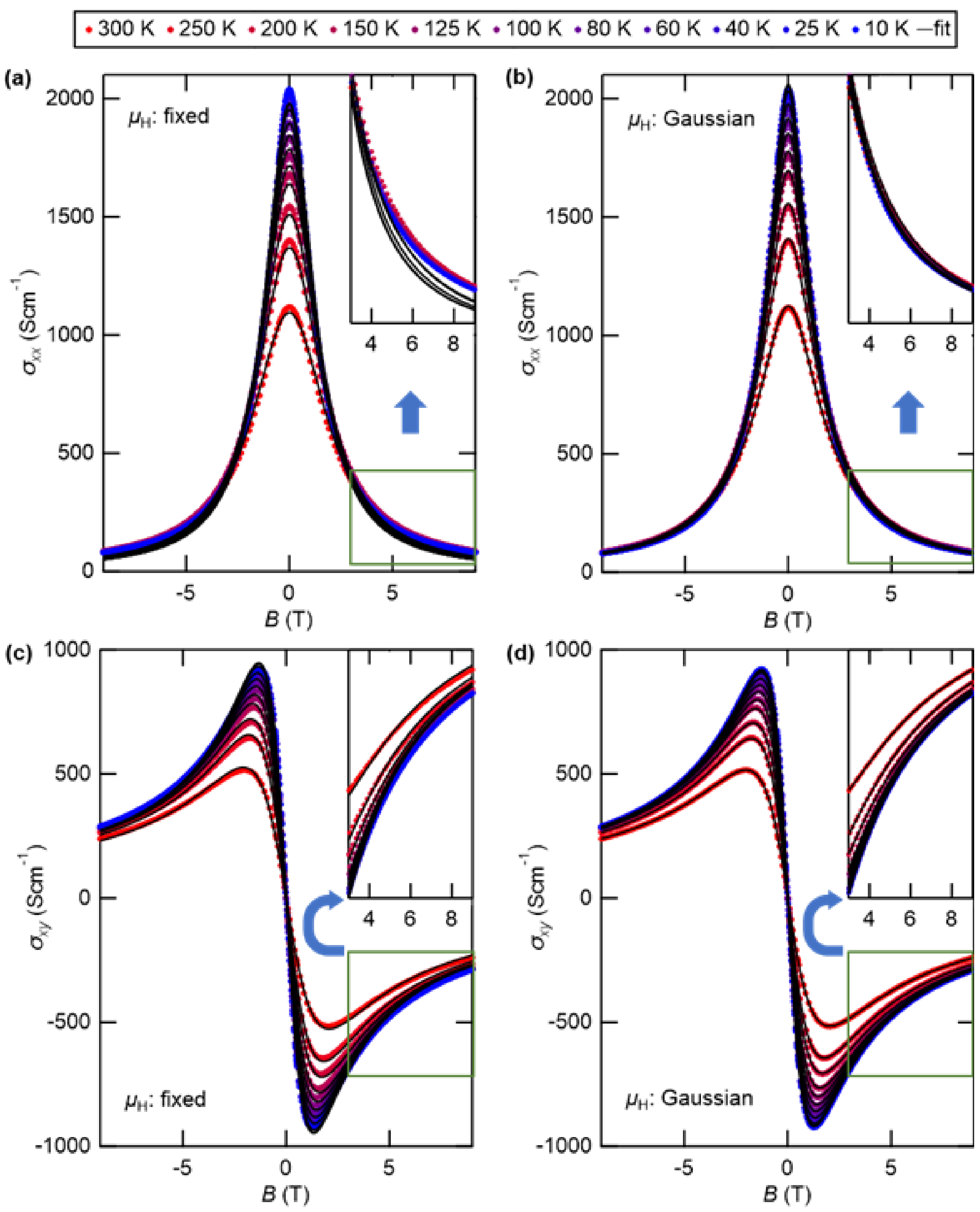

Fig. S5 Magnetic field dependences of $\sigma_{xx}$ and $\sigma_{xy}$ at 10-300 K. (a), (b), (c), and (d) also show the fittings of Eqs. (S27), (S28), (S31), and (S32), respectively. The insets are the enlargements of the high magnetic field regions.

*Contact author: k.kuga@hirosaki-u.ac.jp

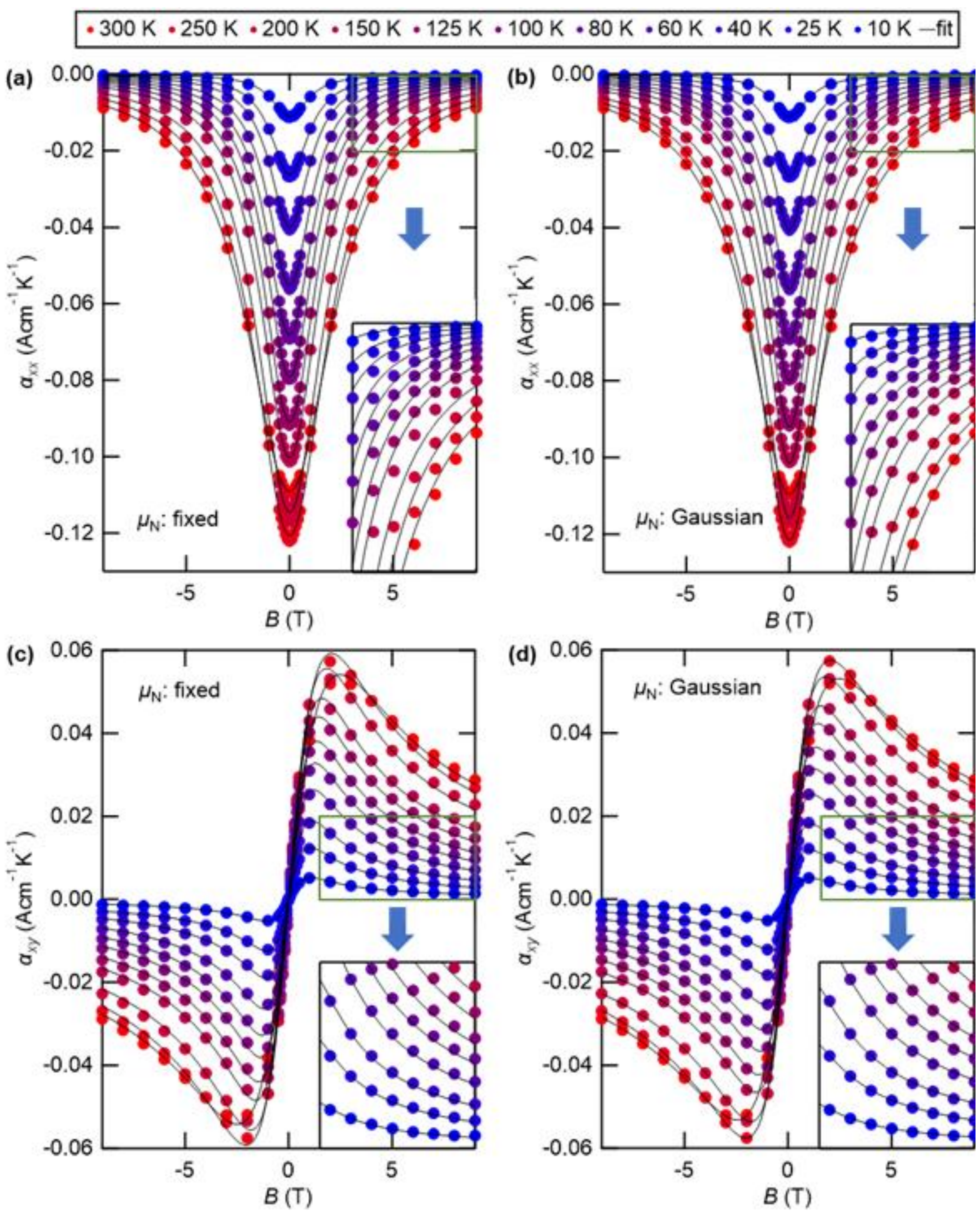

Fig. S6 Magnetic field dependences of $\alpha_{xx}$ and $\alpha_{xy}$ at 10-300 K. (a), (b), (c), and (d) also show the fittings using Eqs. (S29), (S30), (S33), and (S34), respectively. The insets are the enlargements of the high magnetic field regions.

*Contact author: k.kuga@hirosaki-u.ac.jp

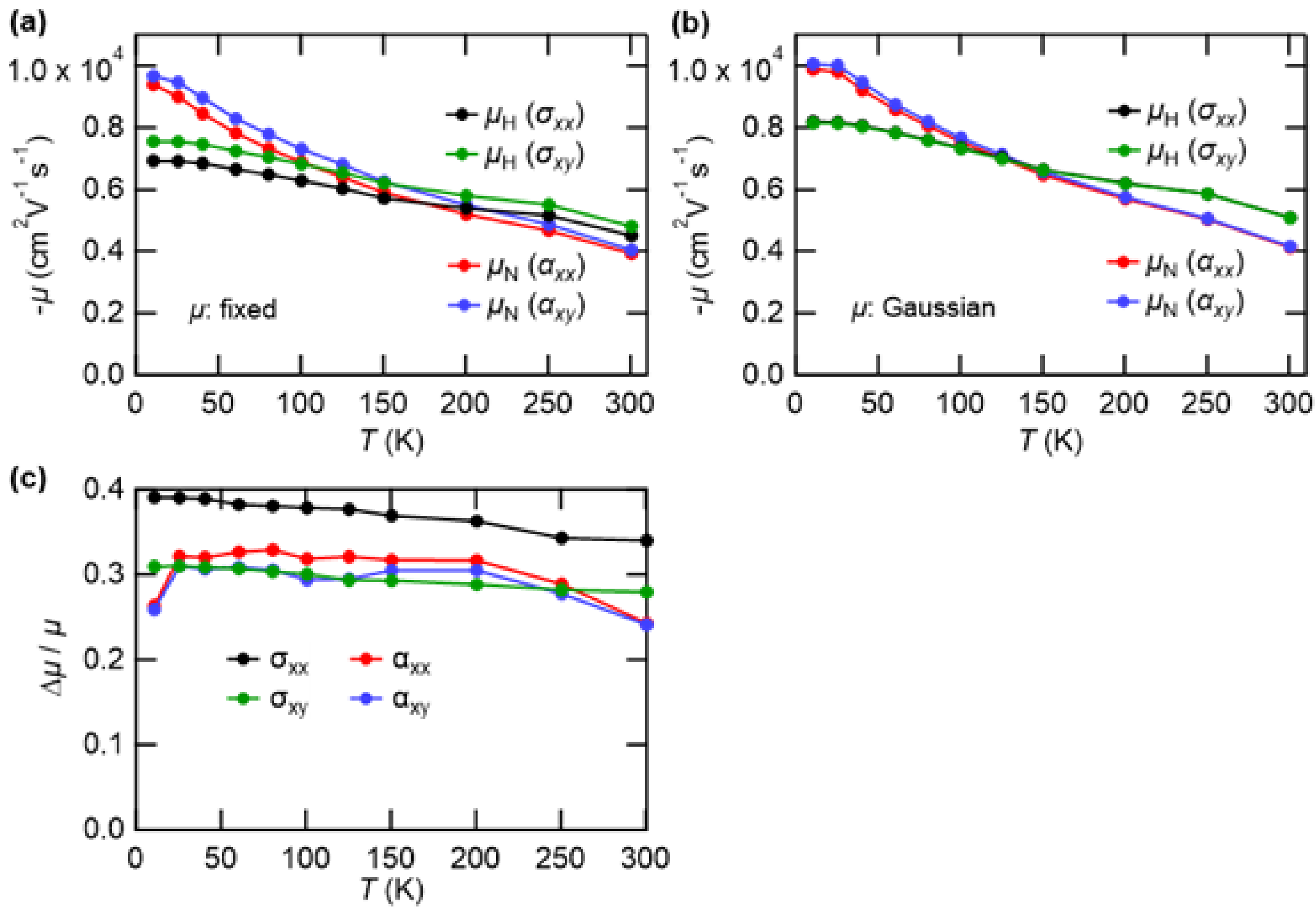

Fig. S7 Temperature dependences of $\mu_{\mathrm{H}}$, $\mu_{\mathrm{N}}$, and the width of Gaussian distribution in the mobility divided by the average of the mobility $\Delta\mu/\mu = w/\mu_{\mathrm{H}}$ or $w/\mu_{\mathrm{N}}$ estimated by $\sigma_{xx}$, $\sigma_{xy}$, $\alpha_{xx}$, and $\alpha_{xy}$. (a) and (b) show $\mu_{\mathrm{H}}$ and $\mu_{\mathrm{N}}$ based on Eqs. (S27-S30) and Eqs. (S31-S34), respectively. (b) is the same as Fig. 4 (a) in the main text.

*Contact author: k.kuga@hirosaki-u.ac.jp

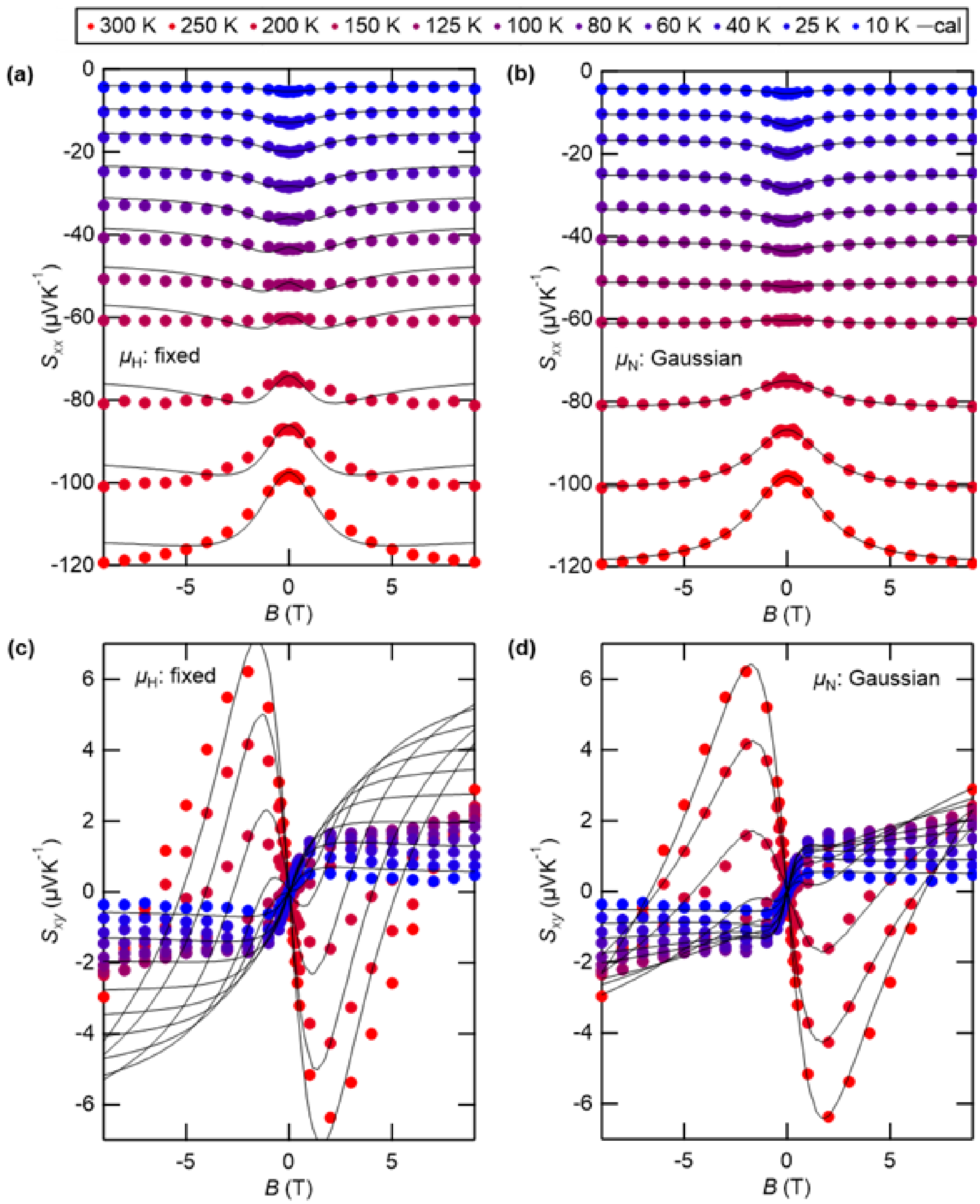

Fig. S8 Magnetic field dependences of $S_{xx}$, $S_{xy}$, and the calculations. The calculations in (a-d) employed experimental $\rho_{xx}$ and $\rho_{xy}$ and the fitting parameters of $S_0$, $\mu_\mathrm{N}$, and $w$ used in Figs. S6 (a-d), respectively.

*Contact author: k.kuga@hirosaki-u.ac.jp

*Contact author: k.kuga@hirosaki-u.ac.jp